\documentclass[useAMS,usenatbib,a4paper]{mn2e}

\usepackage{graphicx}
\usepackage{times}
\usepackage{amsmath}
\usepackage{color}

\def\ipmu{ Kavli Institute for the Physics and Mathematics of the Universe (Kavli IPMU, WPI), University of Tokyo, Chiba 277-8583, Japan}
\def\rceu{ Research Center for the Early Universe, University of Tokyo, 7-3-1 Hongo, Bunkyo-ku, Tokyo 113-0033, Japan}
\def\utok{ Department of Physics, University of Tokyo, 7-3-1 Hongo, Bunkyo-ku, Tokyo 113-0033, Japan}
\def\dlat{ Department of Liberal Arts, Tokyo University of Technology, 5-23-22 Nishikamata, Ota-ku, Tokyo 114-8650, Japan}
\def\ohio{ Department of Astronomy, The Ohio State University, 140 West 18th Avenue, Columbus, OH 43210, USA}
\def\prince{ Princeton University Observatory, Peyton Hall, Princeton, NJ 08544, USA}
\def\ufrgs{ Instituto de F\'isica, UFRGS, CP 15051, Porto Alegre, RS 91501\-970, Brazil} 
\def\linea{Laborat\'orio Interinstitucional de e-Astronomia-LIneA, Rua General Jos\'e Cristino 77, Rio de Janeiro, RJ 20921-400, Brazil}
\def\onac{Observat\'orio Nacional, Rua General Jos\'e Cristino 77, Rio de Janeiro, RJ 20921-400, Brazil}
\def\nara{Department of Physics, Nara National College of Technology, Yamatokohriyama, Nara 639-1080, Japan} 
\def\ccapp{Center for Cosmology and AstroParticle Physics (CCAPP), The Ohio State University, 191 W. Woodruff Ave., Columbus, OH 43210, USA} 
\def\ucdav{Department of Physics, University of California, Davis, CA 95616, USA}
\def\naoj{Optical and Infrared Astronomy Division, National Astronomical Observatory of Japan, 2-21-1, Osawa, Mitaka, Tokyo 181-8588, Japan}
\def\doat{Department of Astronomy, University of Tokyo, 7-3-1 Hongo, Bunkyo-ku, Tokyo 113-0033, Japan} 
\def\phil{Department of Astronomy and Astrophysics, The Pennsylvania State University, University Park, PA 16802 }
\def\igcp{Institute for Gravitation and the Cosmos, The Pennsylvania State University, University Park, PA 16802 }
\def\iacs{Instituto de Astrofísica de Canarias (IAC), E-38200 La Laguna, Tenerife, Spain }
\def\ulls{Universidad de La Laguna (ULL), Dept. Astrofísica, E-38206 La Laguna, Tenerife, Spain }
\def\utah{Department of Physics and Astronomy, University of Utah, 115 S. 1400 E., Salt Lake City, UT 84112, USA }
\def\anna{Physics Department, United States Naval Academy, Annapolis, MD 21403, USA}
\def\epfl{Ecole Polytechnique Fédérale de Lausanne  (EPFL), Observatoire de Sauverny, CH-1290 Versoix}
\def\uchic{The University of Chicago, 5640 South Ellis Avenue, Chicago, IL 60637, USA}
\def\enri{The Enrico Fermi Institute,  5640 South Ellis Avenue, Chicago, IL 60637, USA}
\def\lam{ Aix Marseille Universit\'e, CNRS, LAM (Laboratoire d'Astrophysique de Marseille) UMR 7326, 13388, Marseille, France}
\def\sdsu{Department of Astronomy, San Diego State University, San Diego, CA 92182, USA}

\title[BOSS quasar lens survey]{{The SDSS-III BOSS quasar lens survey:
discovery of thirteen gravitationally lensed quasars}}

\author[More et al.]{ 
\newauthor{
  Anupreeta More$^{1}$, 
  Masamune Oguri$^{1,2,3}$,
  Issha Kayo$^{4}$, 
  Joel Zinn$^{5,6}$, 
}
 \newauthor{
  Michael A. Strauss$^{6}$, 
  Basilio X. Santiago$^{7,8}$, 
  Ana M. Mosquera$^{5,9}$, 
  Naohisa Inada$^{10}$,
}
  \newauthor{
  Christopher S. Kochanek$^{5,11}$, 
  Cristian E. Rusu$^{12,13,14}$, 
  Joel R. Brownstein$^{15}$,
}
 \newauthor {
  Luiz N. da Costa$^{8,16}$, 
  Jean-Paul Kneib$^{17,18}$,
  Marcio A. G. Maia$^{8,16}$,
  Robert M. Quimby$^{1,19}$, 
}
  \newauthor{
  Donald P. Schneider$^{20,21}$, 
  Alina Streblyanska$^{22,23}$, 
  Donald G. York$^{24,25}$
  }

\medskip\\

$^{1}$  \ipmu   \\
$^{2}$  \rceu   \\
$^{3}$  \utok   \\
$^{4}$  \dlat   \\
$^{5}$  \ohio   \\
$^{6}$  \prince \\
$^{7}$  \ufrgs  \\
$^{8}$  \linea  \\
$^{9}$ \anna   \\
$^{10}$  \nara   \\
$^{11}$ \ccapp  \\
$^{12}$ \ucdav  \\
$^{13}$ \naoj   \\
$^{14}$ \doat   \\
$^{15}$ \utah   \\
$^{16}$ \onac   \\
$^{17}$ \epfl   \\
$^{18}$ \lam    \\
$^{19}$ \sdsu   \\ 
$^{20}$ \phil   \\
$^{21}$ \igcp   \\
$^{22}$ \iacs   \\
$^{23}$ \ulls   \\
$^{24}$ \uchic   \\
$^{25}$ \enri   \\
}

\begin{document}


\maketitle

\label{firstpage}

\begin{abstract}
We report the discovery of 13 confirmed two-image quasar lenses from a
systematic search for gravitationally lensed quasars in the SDSS-III
Baryon Oscillation Spectroscopic Survey (BOSS). We adopted a methodology
similar to that used in the SDSS Quasar Lens Search (SQLS).  In addition
to the confirmed lenses, we report 11 quasar pairs with small angular
separations ($\la 2''$) confirmed from our spectroscopy, which are
either projected pairs, physical binaries, or possibly quasar lens
systems whose lens galaxies have not yet been detected.  The newly
discovered quasar lens system, SDSS~J1452+4224 at $z_s\approx 4.8$ is
one of the highest redshift multiply imaged quasars found to date.
Furthermore, we have over 50 good lens candidates yet to be followed up.
Owing to the heterogeneous selection of BOSS quasars, the lens sample
presented here does not have a well-defined selection function. 
\end{abstract}


\section{Introduction}

Gravitationally lensed quasars provide a unique tool to study the
Universe. In particular, the time-variable nature of quasars enable us
to measure the arrival time difference between quasar multiple images,
which may be a powerful probe of cosmology
\citep[e.g.,][]{Refsdal1964,Schechter1997,Treu2013}. Applications of
gravitationally lensed quasars such as time delay cosmography
\citep[e.g.,][]{Coles2008,Suyu2010b}, constraints on the quasar
luminosity function \citep[e.g.,][]{Comerford2002,Richards2006b},
constraints on dark energy \citep[e.g.,][]{Kochanek1996,Oguri2012}, and
study of the host galaxy and/or black hole properties
\citep[e.g.,][]{Peng2006,Mosquera2013,Rusu2015} are still limited by
small number statistics, indicating the importance of finding new
gravitationally lensed quasars in wide-field surveys.

The SDSS Quasar Lens Search
\citep[SQLS;][]{Oguri2006,Oguri2008,Oguri2012,Inada2008,Inada2010,Inada2012}
considerably advanced the field by discovering nearly 50 new quasar
lenses.  The SQLS was based upon the large sample of spectroscopically
confirmed quasars in the Sloan Digital Sky Survey - I/II
\citep[SDSS;][]{York2000}. Candidates were first selected from the $\sim
100,000$ spectroscopic quasars and their SDSS images were then examined
to identify quasars with extended morphology or with nearby companion
objects of similar colour. The lens candidates were then confirmed with
various facilities to construct a secure sample of gravitationally
lensed quasars. Thanks to the well-studied selection function
\citep{Oguri2006}, a sub-sample of the SQLS quasar lenses was used to
place statistical limits on dark energy and the evolution of massive
galaxies that act as lenses \citep{Oguri2012}. 
 
The Baryon Oscillation Spectroscopic Survey \citep[BOSS;][]{Dawson2013}
of the SDSS-III \citep{Eisenstein2011} is a new spectroscopic survey to
measure the expansion rate of the Universe.  In addition to an extensive
spectroscopic survey of $z\sim 0.5$ galaxies, BOSS has obtained spectra
of $\sim 300,000$ quasars \citep{Alam2015} with over 180000 in the
redshift range $2.15<z<4$ in order to detect baryon acoustic
oscillation signatures in the Ly$\alpha$ absorption features of their
spectra.  The large number of BOSS quasars suggests that it should
contain a large number of gravitationally lensed quasars. However, the
lensing rate of the BOSS quasar sample is likely smaller than that of
the SDSS-II quasar sample because BOSS largely targets point-like
quasar candidates in order to increase the survey efficiency. In this
case, lensed quasars where the images are not individually resolved
would appear extended and would tend to be excluded by the BOSS
selection function. On the other hand, the typical BOSS quasar is at a
higher redshift \citep[$z\sim 2-3$][]{Ross2012,Paris2014} than the
typical SDSS-I/II quasars ($z\sim 1-2$) and this higher average redshift
may partly compensate for the lower lensing rate due to the
morphological target selection.  

We thus started a new survey for gravitationally lensed quasars from
BOSS called the BOSS Quasar Lens Survey (BQLS). The basic strategy
follows that of the SQLS in that we select quasar lens candidates by
examining the SDSS images of BOSS quasars and conduct additional
observations of any promising quasar lens candidates. In this paper, we
present 13 confirmed quasar lenses from the BQLS. In addition, we 
also identify 11 quasar close pairs with not quite identical but similar
spectra. Some of these may be true gravitational lenses but could not be
confirmed due to lack of detection of a lens galaxy or spectra being
slightly different due to effects such as microlensing and dust
extinction.

This paper is organized as follows. In Section~\ref{sec:candsel}, we
describe the selection of our lens candidates. In
Section~\ref{sec:data}, we describe the additional observations
undertaken for a detailed examination of the most promising lens
candidates. The mass modelling of the lenses is described in
Section~\ref{sec:massmod}. We compare the confirmed lenses to the SQLS lens
sample and present our general findings in Section~\ref{sec:disc}.
Section~\ref{sec:conc} presents our summary and conclusions. 

\section{Candidate selection}
\label{sec:candsel}

We selected candidate quasar lens systems from the imaging and
spectroscopic data of SDSS-III/BOSS.  The SDSS uses a dedicated
2.5-meter telescope at Apache Point Observatory in New Mexico
\citep{Gunn2006}. The SDSS imaging data uses five
($ugriz$) broadband filters and has a limiting magnitude of $r\sim 22.5$,
which were used to select various targets including quasars for
spectroscopic observations \citep{Ross2012}. The SDSS-III/BOSS survey
uses an upgraded multi-object fiber spectrograph which can
simultaneously take spectra of 1000 objects with a resolving power
$R\sim 2000$ and a wavelength coverage of $3600\AA<\lambda<10400\AA$
\citep{Smee2013}. 

We largely followed the procedures established in the SQLS for the lens
candidate selection process. In the SQLS, however, there were also image
separation and flux ratio limits for constructing a complete lens sample
for statistical studies. In SDSS-III, constructing a well-defined
statistical sample is much more challenging because of the complex
nature of the quasar target selection \citep{Ross2012}.
Thus, we first adopted a simple approach of applying the SQLS-like
 selection criteria and then choosing the best lens candidates by
 visually inspecting the SDSS images of all of the resulting lens candidates. 
 Below we describe these procedures in more detail. 

The selection method for the underlying SDSS-III/BOSS spectroscopic
quasar sample is presented in \citet{Ross2012} and \citet{Paris2014}.
Currently all of the BOSS data are public, and the quasar catalogue we used
is available at
{http://www.sdss.org/dr12/algorithms/boss-dr12-quasar-catalog/}.
From the spectroscopic quasar catalog, we selected quasar lens candidates
using a method similar to SQLS, in which morphological and
colour criteria were used to select lens candidates from the object
catalogue. The full details of the SQLS selection algorithm are
presented in \citet{Oguri2006} and \citet{Inada2008}. While the
SQLS selection criteria were optimized for selecting low-redshift ($z<2.2$)
quasar lens candidates, an extension of the selection algorithm to higher
redshifts is discussed in \citet{Inada2009}. In this paper, we adopted a
modified version of these morphological and colour selection criteria as
described below. 

The morphological selection approach identified spectroscopic quasars
that are poorly fit by the imaging Point Spread Function (PSF).  Even
though BOSS targeted point-like quasars, there are extended objects with
a spectroscopic classification of a quasar in BOSS which were originally
identified as galaxies and targeted as part of the galaxy follow-up
programs or ancillary programs targeting AGNs.  Therefore, we applied
the morphological selection to find small image separation lensed
quasars which appear as extended quasar-like sources. Specifically, we
used the following object parameters; {\tt type}, which
describes the classification of objects as stars (${\tt
  type}=6$) or galaxies (${\tt type}=3$); {\tt lnlStar}, which is the
logarithm of the likelihood of objects being fitted by the PSF of the
field; {\tt lnlDeV}, which is the logarithm of the likelihood of
objects being fitted by the de Vaucouleurs profile. The parameters {\tt
  lnlStar} and {\tt lnlDeV} are available for each broad band. For
quasars at $0.7<z<3$ we selected lens candidates whose $gri$-band
object parameters satisfy ${\tt lnlStar} < -1$ for ${\tt type}=3$ and
${\tt lnlDeV} < -15$ and ${\tt lnlStar} - {\tt lnlDeV} < -5$ for ${\tt
  type}=6$. For quasars at $z>3$, we used the same criteria as above
but only for the $r$ and $i$-bands. There were
$\sim 3000$ candidates selected by these criteria.

For a given spectroscopic quasar, the colour selection method searches
for nearby objects with similar colours out to a maximum image
separation of $\Delta\theta=30''$. For quasars at $0.7<z<3.4$ we used
all 5 bands and select objects with colour differences smaller
than $0.1$ for $g-r$ and $r-i$ and smaller than $0.2$ for $u-g$ and
$i-z$, where the colours were computed using PSF magnitudes. For quasars at
$z>3.4$, we used only the $griz$-bands and similar
thresholds for the colour differences of $0.1$ for $r-i$ and $0.2$ for $g-r$
and $i-z$. We also required that the $i$-band magnitude difference of
the two objects was smaller than $1$. When the angular separation of
the two objects were small ($<3\farcs5$) we used relaxed colour criteria
as the SDSS photometry might not be reliable
\citep[see][]{Oguri2006}. This colour selection method led to a sample of
$\sim 2000$ candidates.  

We then visually inspected all the morphological and colour selected lens
candidates. The visual inspection was performed by one of the authors
(MO) to define ``good'' lens candidates that have relatively high chance
of being true lens systems. Here the goodness of the candidates was
judged by various factors including the presence of multiple components
with quasar-like colours or nearby galaxies that might act as lenses.
Most of the false positives removed in the visual inspection are either
single quasars with contamination from bright nearby objects and obvious
quasar-galaxy superpositions. Large-separation ($\ga 5''$) quasar lenses
are much less common, and hence large-separation lens candidates are not
regarded as good unless putative lens galaxies or clusters are seen in
between the candidate quasar images. We finally selected $\sim 80$ good
candidates and an additional $\sim 70$ possible lens candidates from the
morphological selection, and $\sim 75$  good candidates and $\sim 65$
possible lens candidates from the colour selection. There is an overlap
of $\sim$20 good candidates between the morphology and colour selected
samples.  The number of lens candidates, requiring observational
follow-up studies was reduced by a factor of $\sim 20$ through the visual
inspection. This suggests that our approach is an efficient approach to
find the most promising lens candidates. Nevertheless, it is possible
that we missed some true gravitational lens systems in the course of the
visual inspection. For example, we are likely to miss quasars that are
fainter and do not visually look star-like which includes some small
image-separation lenses and possibly quadruply imaged systems. We may
also have missed systems with atypical image configurations, odd colours or highly
anomalous flux ratios.

We note that \citet{Dahle2013,Dahle2015} discovered a gravitationally
lensed sextuple quasar with a maximum image separation of $15\farcs1$,
SDSS~J2222+2745, from the SDSS-III data. They searched for strong lens
candidates by visually inspecting photometrically identified clusters
of galaxies. Even though this lens system meets our image separation
criteria, none of the lensed quasars from SDSS~J2222+2745 were
selected for spectroscopy by BOSS so this source is not in our
parent sample. 

Some gravitationally lensed quasars have been discovered by identifying
quasar spectral features superposed on the spectra of lens galaxies
\citep[e.g.,][]{Johnston2003,Bolton2005,McGreer2010} although this
technique of identifying compound objects spectroscopically has
primarily discovered many examples of lensed star-forming galaxies
\citep[e.g.,][]{Bolton2006}. As a complementary approach to the SQLS
selection technique, we also searched for lensed quasars using this
technique.  We identified instances of quasar-galaxy superpositions in
the BOSS spectra with quasars at $z<5.5$ and galaxies with redshifts
lower than the quasars (J. Zinn et al. in prep.). The initial
sample had 17 good candidates out of which 3 systems were followed-up
spectroscopically as part of the BQLS sample. Currently, this search
has produced $\sim$100 candidates with a ``good'' visual quality flag
and will be presented in Zinn et al. (in prep.).

The BOSS DR12 catalogue has some known lenses, several of which are in
the SQLS. This known sample was excluded prior to the visual inspection
of the candidates.  In total, we identified about 250 new lens
candidates for further study.  In the next section, we present initial
follow-up results of the most promising quasar lens candidates. 

\section{Data: observation, reduction and analysis}
\label{sec:data}

Our follow-up observations consist of imaging and spectroscopic
observations. We first attempt to obtain follow-up images for the good
lens candidates, and conduct follow-up spectroscopic observations only
when the follow-up images show evidence for gravitational lensing.
However, it was not possible to follow this strategy strictly due to
observing constraints. In some cases we obtained follow-up spectra before
we conducted imaging follow-up observations. Among these systems, those
with clear differences in the spectra were thus not observed with
imaging. Since the spectroscopic observations are very informative for
confirming or ruling out the lensing hypotheses of the candidates, below
we first present the complete spectroscopic follow-up results for the
best 55 candidates, and then present imaging follow-up results for
confirmed gravitational lens systems. The full imaging follow-up results
will be presented in an upcoming paper. 

\subsection{Spectroscopy}
\label{sec:sp}

Table~\ref{tab:spec} summarizes our spectroscopic
observations. We obtained spectra of five candidates with the 
Low Resolution Imaging Spectrometer \citep[LRIS;][]{Oke1995} 
on the Keck telescope on 2013 September 7. We used the $1\farcs0$ slit,
the 400/3400 grism for the blue channel and the 400/8500 grating for the
red channel. The spectral resolution was $R\sim 600$ in the blue
channel and $R\sim 1300$ in the red channel. The
red channel was binned to give $0\farcs27$~pixel$^{-1}$. The unbinned blue
channel has $0\farcs135$~pixel$^{-1}$.  We also obtained spectra of 50 
candidates with the Faint Object Camera and Spectrograph
\citep[FOCAS;][]{Kashikawa2002} on the Subaru telescope on 2014 
May 2, 2015 Feb 19, and 2015 June 19. In all of the FOCAS observations 
we used the $1\farcs0$ slit, but adopted two different configurations 
depending on the quasar redshifts; one is the grism and filter set of
300B/L600 to cover the wavelength range of $3700-6000$~\AA, and the
other is 300B/SY47 $4700-9100$~\AA. Both datasets have a spectral
resolution of $R\sim 400-500$. The data were binned $2\times 2$ on the
detector, providing a spatial resolution of $0\farcs21$~pixel$^{-1}$. In
all of the observations, the long-slit was aligned to observe both 
putative quasar images simultaneously. The data were reduced using
 standard IRAF\footnote{IRAF is distributed by the National Optical Astronomy
  Observatories, which are operated by the Association of Universities
  for Research in Astronomy, Inc., under cooperative agreement with
  the National Science Foundation.} tasks.  

\begin{table*}
\begin{center}
\caption{\label{tab:spec}Summary of spectroscopic 
  observations. }
\begin{tabular}{ccccclrl}
\hline
Object ID & RA (J2000)  & Dec (J2000)  & Selection & $z_{\rm QSO}$ & Observation & Exp. [s] & Result\\ 
\hline
SDSS~J0033+2015 & 00:33:37.59 &   +20:15:38.2 & C  & 2.701 & K, 2013 Sep (400/3400+400/8500) & 600 & quasar+star  \\   
SDSS~J0035+2659 & 00:35:31.98 &   +26:59:59.9 & M  & 2.294 & S, 2015 Jun (300B/L600) & 1500 & different SED\\
SDSS~J0114+0722 & 01:14:38.38 &   +07:22:28.5 & MC & 1.828 & K, 2013 Sep (400/3400+400/8500) & 600 & quasar lens ($z_l=0.408$)\\
SDSS~J0139+1908 & 01:39:40.69 &   +19:08:40.7 & MC & 3.095 & K, 2013 Sep (400/3400+400/8500) & 600 & different SED\\
SDSS~J0206+0440 & 02:06:49.50 &   +04:40:19.0 & C  & 2.396 & K, 2013 Sep (400/3400+400/8500) & 600 & different SED\\
SDSS~J0213-0421 & 02:13:22.86 & $-$04:21:34.3 & C  & 1.910 & S, 2015 Feb (300B/L600) & 900 & quasar pair $z=1.911\&0.992$\\ 
SDSS~J0256+0153 & 02:56:40.76 &   +01:53:29.3 & S  & 2.600 & K, 2013 Sep (400/3400+400/8500) & 600 & quasar lens ($z_l=0.603$)\\
SDSS~J0737+4825 & 07:37:08.67 &   +48:25:51.1 & M  & 2.892 & S, 2015 Feb (300B/L600) & 1080 & quasar lens\\
SDSS~J0757+2150 & 07:57:20.54 &   +21:50:07.7 & MC & 2.128 & S, 2014 May (300B/L600) & 720 & quasar+galaxy at $z=0.118$\\
SDSS~J0818+0601 & 08:18:30.46 &   +06:01:38.0 & M  & 2.381 & S, 2015 Feb (300B/L600) & 900 & quasar pair $z=2.359\&2.361$\\
SDSS~J0821+0735 & 08:21:43.36 &   +07:35:45.9 & C  & 2.378 & S, 2015 Feb (300B/L600) & 1080 & quasar pair $z=2.383\&2.387$\\
SDSS~J0821+4542 & 08:21:58.66 &   +45:42:44.4 & M  & 2.066 & S, 2014 May (300B/L600) & 900 & quasar lens ($z_l=0.349$)\\
SDSS~J0826+4248 & 08:26:52.99 &   +42:48:17.8 & MS & 1.942 & S, 2014 May (300B/L600) & 900 & inconclusive\\
SDSS~J0847+1504 & 08:47:21.96 &   +15:04:50.2 & M  & 3.280 & S, 2015 Feb (300B/SY47) & 900 & different SED\\
SDSS~J0921+2854 & 09:21:15.47 &   +28:54:44.3 & MC & 1.410 & S, 2015 Feb (300B/L600) & 720 & quasar lens ($z_l=0.445$$^a$)\\
SDSS~J0928+4332 & 09:28:39.19 &   +43:32:42.4 & MC & 3.698 & S, 2014 May (300B/SY47) & 900 & quasar pair $z=3.694\&2.995$\\
SDSS~J0930+4614 & 09:30:21.16 &   +46:14:22.8 & M  & 2.397 & S, 2015 Feb (300B/L600) & 900 & quasar pair $z=2.393\&2.394$\\
SDSS~J0958+0744 & 09:58:42.24 &   +07:44:23.2 & M  & 2.781 & S, 2014 May (300B/L600) & 900 & quasar+galaxy at $z=0.158$\\
SDSS~J1001+0156 & 10:01:40.95 &   +01:56:43.1 & M  & 2.202 & S, 2015 Feb (300B/L600) & 1080 & different SED\\
SDSS~J1043+4320 & 10:43:24.87 &   +43:20:49.4 & M  & 2.234 & S, 2015 Jun (300B/L600) & 900 & quasar pair $z=2.245\&2.225$\\
SDSS~J1108+4726 & 11:08:28.25 &   +47:26:21.6 & C  & 3.627 & S, 2014 May (300B/SY47) & 900 & quasar+galaxy at $z=0.385$\\
SDSS~J1124+5710 & 11:24:55.24 &   +57:10:56.5 & C  & 2.312 & S, 2015 Feb (300B/L600) & 720 & quasar pair $z=2.309\&2.315$\\
SDSS~J1125+5020 & 11:25:42.60 &   +50:20:35.5 & C  & 1.066 & S, 2015 Jun (300B/L600) & 900 & quasar+galaxy at $z=0.385$\\
SDSS~J1138+5254 & 11:38:38.54 &   +52:54:18.2 & M  & 2.758 & S, 2015 Jun (300B/L600) & 1200 & different SED\\
SDSS~J1139-0014 & 11:39:28.49 & $-$00:14:18.1 & C  & 3.084 & S, 2014 May (300B/L600) & 720 & quasar+star\\
SDSS~J1152+2235 & 11:52:10.58 &   +22:35:19.2 & C  & 2.902 & S, 2014 May (300B/L600) & 720 & quasar+star\\
SDSS~J1248+6104 & 12:48:59.85 &   +61:04:30.3 & C  & 2.591 & S, 2015 Feb (300B/L600) & 1200 & quasar+star\\
SDSS~J1254+1857 & 12:54:40.37 &   +18:57:12.0 & MC & 1.717 & S, 2015 Feb (300B/L600) & 1200 & quasar lens ($z_l=0.555$$^a$)\\
SDSS~J1309+5617 & 13:09:27.55 &   +56:17:38.9 & C  & 2.505 & S, 2015 Jun (300B/L600) & 1200 & quasar pair $z=2.513\&2.515$\\
SDSS~J1319+5023 & 13:19:26.06 &   +50:23:05.5 & C  & 2.301 & S, 2015 Jun (300B/L600) & 1080 & quasar+galaxy at $z=0.088$\\ 
SDSS~J1322+3038 & 13:22:04.81 &   +30:38:38.2 & M  & 2.233 & S, 2015 Jun (300B/L600) & 1200 & inconclusive\\
SDSS~J1330+3800 & 13:30:07.34 &   +38:00:42.4 & C  & 2.254 & S, 2015 Feb (300B/L600) & 720 & quasar lens\\
SDSS~J1405+1350 & 14:05:56.92 &   +13:50:38.3 & C  & 2.342 & S, 2015 Jun (300B/L600) & 1200 & quasar pair $z=2.345\&2.361$\\
SDSS~J1412+5204 & 14:12:10.18 &   +52:04:23.3 & M  & 2.952 & S, 2015 Feb (300B/L600) & 1200 &  quasar lens\\
SDSS~J1429+2523 & 14:29:38.20 &   +25:23:43.4 & M  & 3.904 & S, 2015 Feb (300B/SY47) & 900 & different SED\\
SDSS~J1442+4055 & 14:42:54.79 &   +40:55:35.6 & C  & 2.575 & S, 2014 May (300B/L600) & 480 & quasar lens\\
SDSS~J1452+4224 & 14:52:11.50 &   +42:24:29.6 & M  & 4.819 & S, 2015 Feb (300B/SY47) & 900 & quasar lens ($z_l=0.382$)\\
SDSS~J1458-0202 & 14:58:47.59 & $-$02:02:05.9 & M  & 1.724 & S, 2014 May (300B/L600) & 1200& quasar lens\\
SDSS~J1508+3037 & 15:08:22.32 &   +30:37:47.2 & M  & 2.449 & S, 2015 Feb (300B/L600) & 1200 & inconclusive\\
SDSS~J1537+3014 & 15:37:34.46 &   +30:14:53.7 & MCS& 1.553 & S, 2014 May (300B/L600) & 900 & quasar lens ($z_l=0.490$$^a$)\\
SDSS~J1548+2830 & 15:48:50.76 &   +28:30:14.3 & C  & 3.227 & S, 2015 Feb (300B/SY47) & 900 & quasar pair $z=3.208\&1.487$\\
SDSS~J1551+3303 & 15:51:37.02 &   +33:03:19.6 & C  & 3.806 & S, 2015 Feb (300B/SY47) & 900 & different SED\\
SDSS~J1552+2401 & 15:52:21.40 &   +24:01:23.1 & MC & 3.674 & S, 2015 Feb (300B/SY47) & 720 & quasar+star\\
SDSS~J1554+2616 & 15:54:11.53 &   +26:16:35.7 & C  & 2.321 & S, 2015 Feb (300B/L600) & 720 & quasar+star\\
SDSS~J1556+1731 & 15:56:23.81 &   +17:31:21.4 & C  & 2.814 & S, 2014 May (300B/L600) & 720 & quasar+galaxy at $z=0.109$\\
SDSS~J1600+3148 & 16:00:28.78 &   +31:48:31.7 & C  & 2.331 & S, 2015 Jun (300B/L600) & 1080 & quasar+star\\
SDSS~J1611+0844 & 16:11:05.64 &   +08:44:35.4 & C  & 4.548 & S, 2015 Feb (300B/SY47) & 900 & different SED\\
SDSS~J1627+5553 & 16:27:16.69 &   +55:53:37.0 & C  & 4.072 & S, 2015 Feb (300B/SY47) & 720 & different SED\\
SDSS~J1712+2516 & 17:12:32.14 &   +25:16:24.6 & M  & 2.604 & S, 2015 Feb (300B/L600) & 1080 & quasar+star\\
SDSS~J1715+2831 & 17:15:17.32 &   +28:31:29.5 & C  & 2.023 & S, 2015 Feb (300B/L600) & 1080 & different SED\\
SDSS~J2146-0047 & 21:46:46.03 & $-$00:47:44.3 & M  & 2.381 & S, 2014 May (300B/L600) & 900 & quasar lens at $z=0.799$\\
SDSS~J2220-0050 & 22:20:41.31 & $-$00:50:15.5 & MC & 2.602 & S, 2015 Jun (300B/L600) & 1080 & quasar+star\\
SDSS~J2238+1820 & 22:38:51.80 &   +18:20:38.5 & MC & 2.074 & S, 2015 Jun (300B/L600) & 720 & quasar+galaxy at $z=0.199$\\
SDSS~J2245+2548 & 22:45:55.75 &   +25:48:35.4 & C  & 2.992 & S, 2015 Jun (300B/L600) & 900 & quasar pair $z=2.995\&2.178$\\
SDSS~J2314-0108 & 23:14:24.50 & $-$01:08:58.7 & C  & 3.455 & S, 2015 Feb (300B/SY47) & 1080 & different SED\\
\hline
\end{tabular}
\flushleft{The `Selection' column indicates the method(s) with `M' for
morphological selection, `C' for colour selection, and `S' for
spectroscopic selection (see text for details). Several candidates were
selected by multiple methods. `Observation' indicates the telescope
(S$-$Subaru or K$-$Keck), the date and setup of the spectroscopic
observations (see also Sec.~\ref{sec:sp}).  `Result' indicates the
conclusion from the follow-up spectroscopy as well as imaging
observations described in Sec.~\ref{sec:im}.  The $z_{\rm QSO}$ are
taken from the BOSS DR12 catalogue which has redshifts corrected after
visual inspection of the spectra. The last column has redshifts from the
follow-up spectra except when marked with $^a$ which are taken from the
BOSS spectrum.} 
\end{center}
\end{table*}

We found that thirteen out of the 55 quasar lens candidates contain two
quasar components with identical redshifts and similarities in the
Spectral Energy Distributions (SED). Together with the analysis of the
imaging observations presented below, we conclude that these 13 systems
are gravitationally lensed quasars.  Our spectroscopy also identified
an additional 11 quasar pairs with either almost identical or slightly
different redshifts. Properties of these 11 quasar pairs will be
discussed in Sec.~\ref{sec:individual} and \ref{sec:qpair}.

\begin{figure*}
\begin{center}
\includegraphics[scale=0.43]{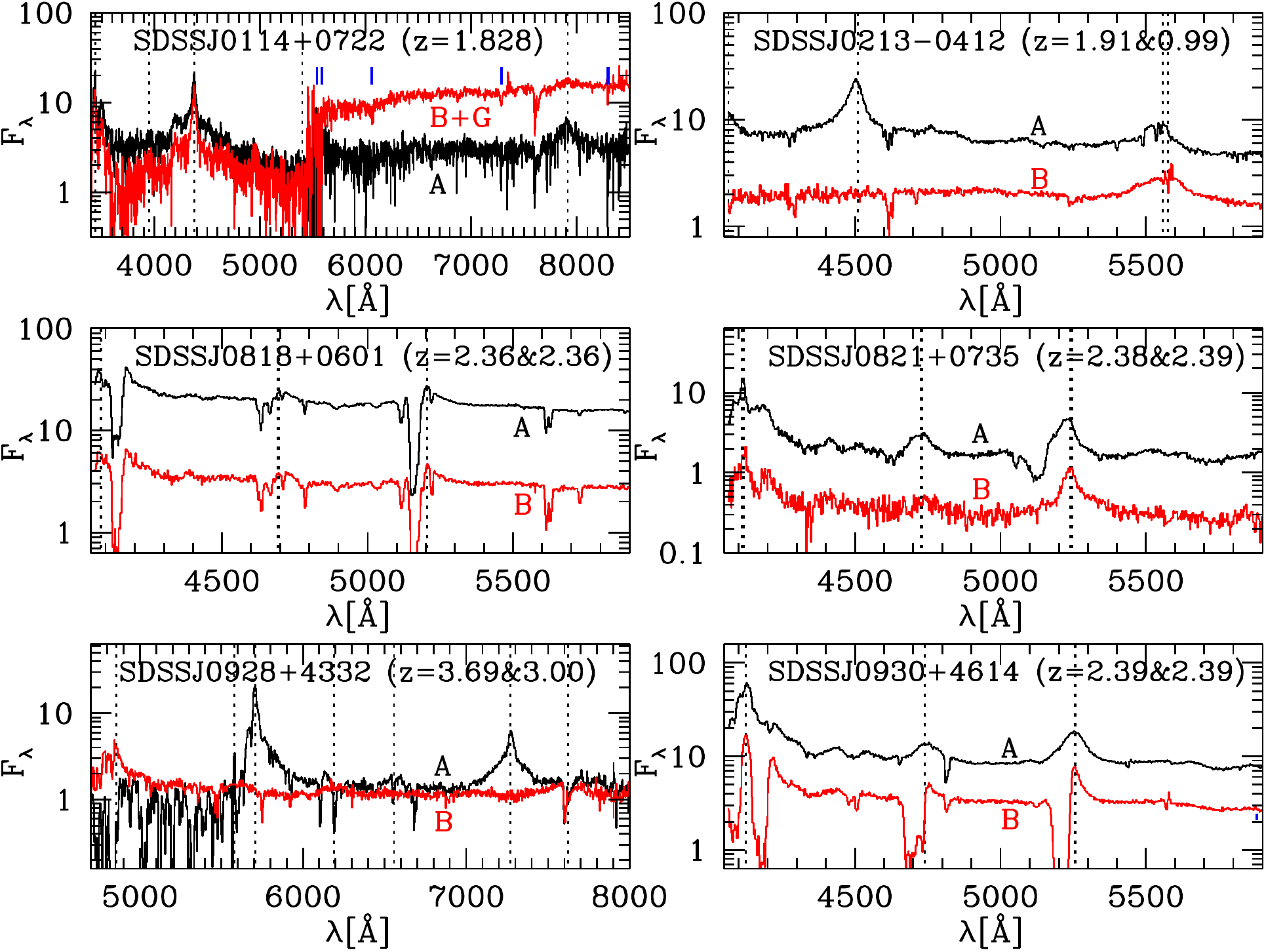}
\includegraphics[scale=0.43]{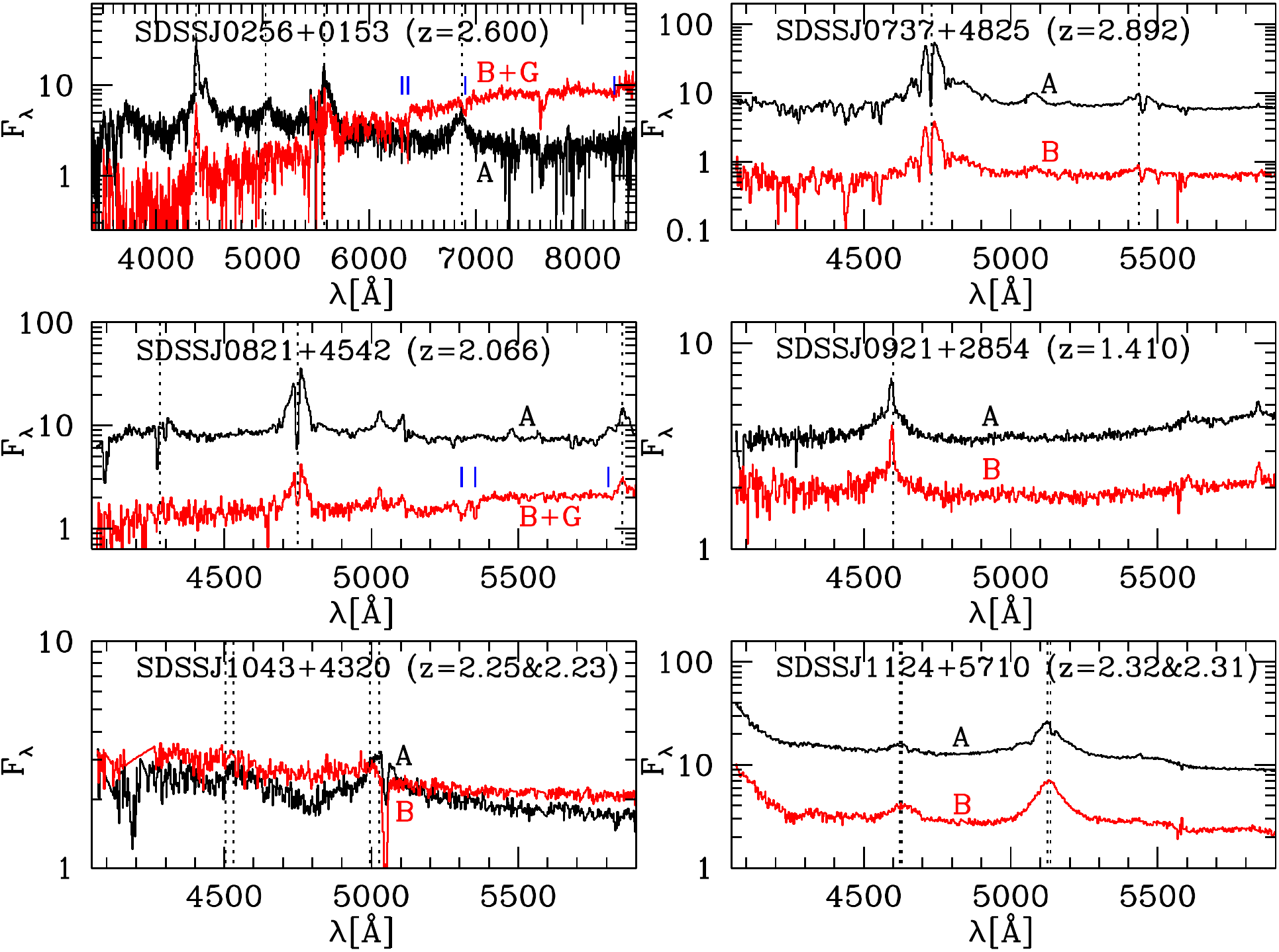}
\includegraphics[scale=0.43]{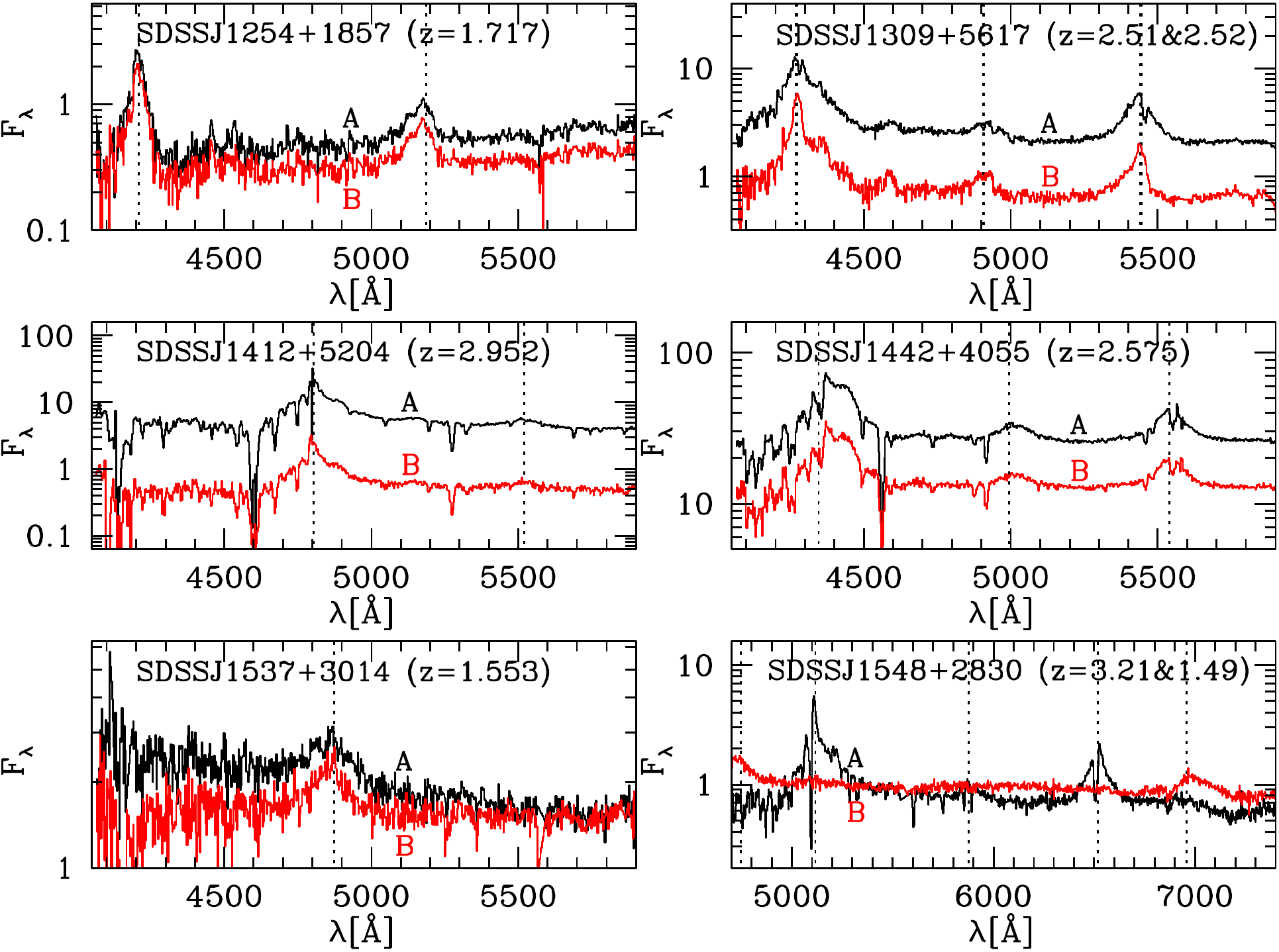}
\includegraphics[scale=0.43]{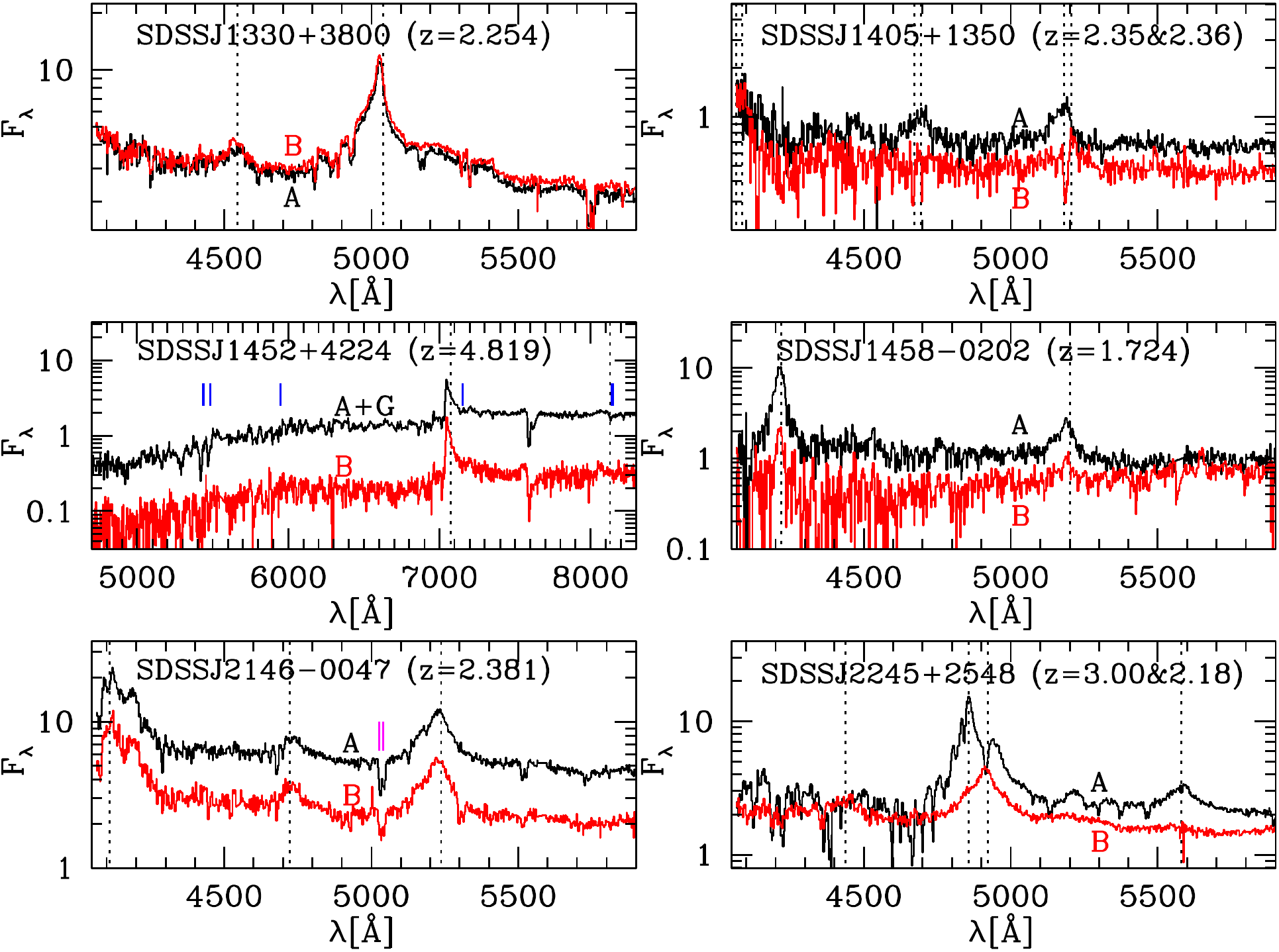}
\caption{Spectra of the 24 confirmed quasar pairs, including 13
confirmed gravitationally lensed quasars (see Table~\ref{tab:spec}). In
each panel, vertical dotted lines show the wavelengths of the quasar
emission lines. Vertical short bars indicate the wavelengths of
absorption features in the spectrum of the lens galaxy, which are shown
only for lens systems for which lens galaxies are sufficiently bright to
be detected in one of the quasar spectra. The redshifts shown in each
panel correspond to the quasar. The flux $F_\lambda$ is in units of
$10^{-17}~{\rm erg\,cm^{-2}s^{-1}\AA^{-1}}$.
\label{fig:spec} }
\end{center}
\end{figure*}

Figure~\ref{fig:spec} presents the spectra for the 24 confirmed
quasar pairs, including the 13 confirmed gravitationally lensed
quasars. For five of the quasar lens systems, the lens galaxies are
bright enough to be clearly detected in one of the quasar spectra. We
successfully measured the lens redshift for these lens systems (see
Table~\ref{tab:spec}). For an additional three quasar systems, the
lens galaxies are bright and their redshifts were derived from the SDSS-III spectra. 
The subsequent spectra do not cover the red
wavelength range where these lens galaxies can be detected easily.

\subsection{Imaging}
\label{sec:im}

\begin{table}
\begin{center}
\caption{\label{tab:imag} Summary of imaging observations
  of the confirmed lenses.}
\begin{tabular}{cccc}
\hline
Object &  Instrument & Observing Date & Exposure \\ 
ID &   &  & (sec) \\ 
\hline
J0114+0722  &  Tek2k  &   2013 Dec 24  &  1440 (V,I)     \\   
J0256+0153  &  Tek2k  &   2013 Dec 24  &  1440 (V,I)     \\
J0737+4825  &  Tek2k  &   2015 Feb 20  &  1440 (V,I)\\
J0821+4542  &  Tek2k  &   2014 Feb 21  &  2160 (V) 1980 (R)\\
J0921+2854  &  Tek2k  &   2015 Feb 20  &  1440 (V) 1080 (I)\\
J1254+1857  &  Tek2k  &   2014 Feb 21  &  3060 (V) 2880 (R)\\
J1330+3800  &  Tek2k  &   2014 Feb 21  &  1440 (V) 2160 (R) \\
J1412+5204  &  Tek2k  &   2015 Feb 20  &  1440 (V,I)\\
J1442+4055  &  OSMOS  &   2013 Apr 08  &  1200 (g) 2100 (r) 900 (i,z) \\
J1452+4224  &  FOCAS  &   2015 Jun 19  &  270 (R) 360 (I)\\
J1458-0202  &  SOI    &   2013 Feb 12  &  420 (g,r,i)   \\
J1537+3014  &  Tek2k  &   2014 Feb 21  &  1440 (V) 1800 (R) \\
J2146-0047  &  Tek2k  &   2012 Sep 17  &  960 (V,I)     \\
\hline
\end{tabular}
Tek2k is on the UH88 telescope, OSMOS is on the Hiltner telescope, FOCAS
is on the Subaru telescope and SOI is at the SOAR telescope.
\end{center}
\end{table}

We imaged a sub-sample of the most promising candidates in $2-3$ bands
with the goal of detecting the lens galaxy.  At minimum, these imaging
data should better resolve the small separation lensed quasar images to
provide the basic constraints required for mass modelling. The majority
of the lenses from the BQLS sample were observed with the Tektronix
2048$\times$2048 CCD camera (Tek2k, with a pixel resolution of
$0\farcs22$) on the UH88 telescope. A few of the candidates were
observed with other telescopes. A summary of these observations is given
in Table~\ref{tab:imag} which includes the exposure times for each
filter. We used the FOCAS instrument with $2\times2$ binning on Subaru
giving a pixel resolution of $0\farcs208$, the SOI instrument (pixel
resolution of $0\farcs15$) on the SOuthern Astrophysical Research (SOAR)
telescope and the OSMOS instrument (pixel resolution of $0\farcs273$) on
the Hiltner telescope of the MDM observatory. We processed the data in
IRAF by following standard procedures. The seeing of our imaging data is
better than 1\arcsec. 

\begin{figure*}
\begin{center}
\includegraphics[scale=1.10]{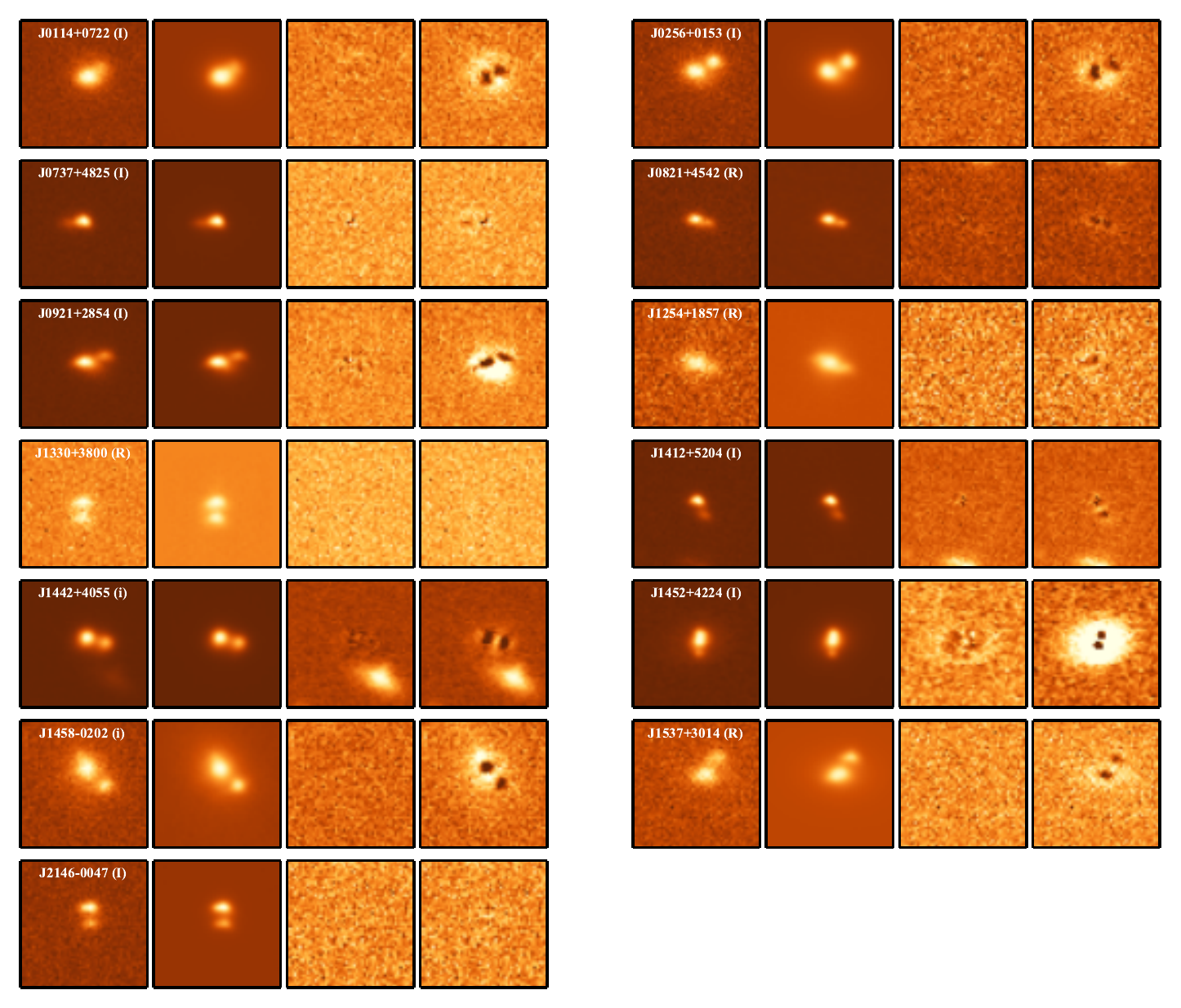}
\caption{ \label{fig:igal} {\sc galfit} modelling results for the quasar
lens sample using the reddest available band as labelled in each case.
For each lens system, the columns from left to right are the data, the
{\sc galfit} model including the lens galaxy and the residuals with and
without the lens galaxy, respectively. The flux scales in the data and
model images match but they are different from the residual images. The
flux scales of both residual images match. All images have the standard
orientation with North up and East on the left. All images are 51
pixels on the side, which is $\sim$11\arcsec for all systems except for
J1442+4055 ($\sim$14\arcsec) and J1458-0202 ($\sim$8\arcsec).}
\end{center}
\end{figure*}

In most cases, the imaging reveals two point-like components with a hint
of extended emission close to one of those components indicating
presence of a lens galaxy. We use {\sc galfit} \citep{Peng2002,Peng2010}
to model each of these components and to measure their fluxes and
positions.  The quasars are assumed to be point sources convolved with
the PSF and the galaxy is modelled with a Sersic profile.  The use of
point sources for the quasar components is reasonable because all the
quasars are at high redshifts $z>1$.  We use the nearby stars as a model
for the PSF. We first fit models to the two quasars without the lens.
Next, we fit models including the lens. Both modelling results are
compared and the latter model is accepted if the $\chi^2$ and the
residuals have improved. In the cases when it was difficult to fit the
Sersic index of the galaxy's light profile, we held it fixed at $n=4$,
since  the de Vaucouleurs profile \citep{deVaucouleurs1948} is a good
model for early-type galaxies. The modelling results for all of the 13
confirmed lenses are presented in Figure~\ref{fig:igal}. For each lens,
we show the reddest available image (left panel) where the contribution
from the lens galaxy is easier to see than in the other bands, the {\sc
galfit} model which includes the lens galaxy (second panel from left)
and the residuals from this model (third panel from left). For
comparison, we also show the residuals from the model without the lens
galaxy in the right-most panel.  

We performed relative photometry of the lenses by comparing them to
nearby stars with known SDSS magnitudes. First, we used the software
from Astrometry.net \citep{Lang2010} to determine the orientations of the images and the
positions of the stars, except for SDSS~J1458-0202 and SDSS~J1452+4224
where the astrometry was determined manually. For every lens in every band available from the
imaging, we measured the fluxes of about a dozen stars using {\sc
Sextractor} \citep{Bertin1996}. The zero points were determined by comparing the SDSS
magnitudes to the fluxes of the stars. Owing to the differences between
the SDSS filters and those used for the follow up observations, we had
to apply appropriate filter conversions to the fluxes.  We adopted the
conversions from \citet{Jester2005} and Lupton (2005, unpublished) which
are available at
{https://www.sdss3.org/dr8/algorithms/sdssUBVRITransform.php}. The zero
points were then used for determining the magnitudes of the lensed
quasars and lens galaxies. The results of the relative astrometry and
photometry are given in Table~\ref{tab:posmag}. The best fit parameters
of the Sersic model assumed for the lens galaxies are reported in
Table~\ref{tab:galsers}.

\begin{table*}
\begin{center}
\caption{\label{tab:posmag} Relative astrometry and photometry of the 13
confirmed lens systems}
\begin{tabular}{cccccc}
\hline
Object ID & Component & Ra & Dec  & Red & Blue \\ 
and filters & & (J2000) & (J2000) & ($I$, $i$ or $R$) &  ($V$, $g$ or$R$) \\
\hline
 SDSS~J0114+0722 & A & 01:14:38.421 & +07:22:28.50 & 20.88$\pm$0.07 & 22.70$\pm$0.17  \\
$I$,$V$          & B & 01:14:38.322 & +07:22:29.35 & 21.20$\pm$0.03 & 22.98$\pm$0.09 \\ 
                 & G & 01:14:38.386 & +07:22:28.60 & 19.30$\pm$0.03 & 20.56$\pm$0.03 \\ 
\hline
 SDSS~J0256+0153 & A & 02:56:40.771 & +01:53:29.28 & 21.20$\pm$0.05 & 22.60$\pm$0.09  \\
$I$,$V$          & B & 02:56:40.653 & +01:53:30.06 & 20.79$\pm$0.01 & 22.33$\pm$0.04 \\ 
                 & G & 02:56:40.740 & +01:53:29.12 & 19.89$\pm$0.04 & 20.50$\pm$0.89 \\ 
\hline
 SDSS~J0737+4825 & A & 07:37:08.677 & +48:25:51.45 & 18.28$\pm$0.00 & 19.26$\pm$0.00  \\
$I$,$V$          & B & 07:37:08.829 & +48:25:51.26 & 20.58$\pm$0.07 & 22.03$\pm$0.14 \\ 
                 & G & 07:37:08.782 & +48:25:51.32 & 19.87$\pm$0.10 & 22.11$\pm$0.15 \\ 
\hline
 SDSS~J0821+4542 & A & 08:21:58.706 & +45:42:44.40 & 20.76$\pm$0.14 & 20.01$\pm$0.02  \\
$R$,$V$          & B & 08:21:58.582 & +45:42:44.07 & 20.86$\pm$0.02 & 21.54$\pm$0.07 \\ 
                 & G & 08:21:58.678 & +45:42:44.38 & 20.01$\pm$0.06 & -- \\ 
\hline
 SDSS~J0921+2854 & A & 09:21:15.494 & +28:54:44.57 & 18.03$\pm$0.01 & 20.36$\pm$0.01  \\
$I$,$V$          & B & 09:21:15.355 & +28:54:45.10 & 18.93$\pm$0.01 & 20.93$\pm$0.01 \\ 
                 & G & 09:21:15.433 & +28:54:44.35 & 18.28$\pm$0.01 & 20.28$\pm$0.08 \\ 
\hline
 SDSS~J1254+1857 & A & 12:54:40.445 & +18:57:12.33 & 21.98$\pm$0.18 & --  \\
$R$              & B & 12:54:40.295 & +18:57:11.44 & 21.74$\pm$0.07 & -- \\ 
                 & G & 12:54:40.385 & +18:57:11.89 & 20.02$\pm$0.11 & -- \\ 
\hline
 SDSS~J1330+3800 & A & 13:30:07.330 & +38:00:43.82 & 20.16$\pm$0.11 & 19.52$\pm$0.02  \\
$R$,$V$          & B & 13:30:07.343 & +38:00:42.39 & 20.03$\pm$0.04 & 19.69$\pm$0.02 \\ 
                 & G & 13:30:07.375 & +38:00:43.64 & 19.74$\pm$0.15 & -- \\ 
\hline
 SDSS~J1412+5204 & A & 14:12:10.158 & +52:04:23.50 & 18.91$\pm$0.01 & 19.85$\pm$0.00  \\
$I$,$V$          & B & 14:12:10.080 & +52:04:22.16 & 20.98$\pm$0.09 & 22.00$\pm$0.02 \\ 
                 & G & 14:12:10.114 & +52:04:22.46 & 20.16$\pm$0.06 & -- \\ 
\hline
 SDSS~J1442+4055 & A & 14:42:54.795 & +40:55:35.74 & 17.64$\pm$0.00 & 18.14$\pm$0.00  \\
$i$,$g$          & B & 14:42:54.614 & +40:55:35.18 & 18.28$\pm$0.01 & 18.83$\pm$0.00 \\ 
                 & G & 14:42:54.696 & +40:55:35.38 & 18.83$\pm$0.05 & 20.07$\pm$0.10 \\ 
\hline
 SDSS~J1452+4224 & A & 14:52:11.500 & +42:24:29.02 & 20.08$\pm$0.02 & 21.05$\pm$0.02  \\
$I$,$R$          & B & 14:52:11.490 & +42:24:30.61 & 20.91$\pm$0.02 & 21.71$\pm$0.02 \\ 
                 & G & 14:52:11.490 & +42:24:29.62 & 18.15$\pm$0.02 & 18.68$\pm$0.06 \\ 
\hline
 SDSS~J1458-0202 & A & 14:58:47.580 & $-$02:02:05.16 & 21.50$\pm$0.02 & 24.50$\pm$0.10  \\
$i$,$g$          & B & 14:58:47.670 & $-$02:02:03.49 & 23.16$\pm$0.15 & 22.76$\pm$0.02 \\
                 & G & 14:58:47.650 & $-$02:02:04.19 & 18.96$\pm$0.07 & 21.51$\pm$0.08 \\ 
\hline
 SDSS~J1537+3014 & A & 15:37:34.484 & +30:14:53.70 & 22.02$\pm$0.15 & 21.95$\pm$0.10  \\
$R$,$V$          & B & 15:37:34.375 & +30:14:55.49 & 21.14$\pm$0.03 & 21.97$\pm$0.18 \\ 
                 & G & 15:37:34.442 & +30:14:54.12 & 18.95$\pm$0.29 & 20.84$\pm$0.06 \\ 
\hline                                
 SDSS~J2146-0047 & A & 21:46:46.025 & $-$00:47:44.10 & 19.74$\pm$0.02 & 20.29$\pm$0.04  \\
$I$,$V$          & B & 21:46:46.018 & $-$00:47:45.48 & 20.54$\pm$0.05 & 21.57$\pm$0.19 \\ 
                 & G & 21:46:46.020 & $-$00:47:44.82 & 21.37$\pm$0.32 & 21.29$\pm$0.31 \\ 
\hline

\end{tabular}
\end{center}
\end{table*}

\begin{table}
\begin{center}
\caption{\label{tab:galsers} Sersic parameters of the lens
  galaxies.}
\begin{tabular}{ccccc}
\hline
Object & $r_e$ & $n$ & $e$ & $\theta_e$ \\ 
ID & (\arcsec) &  &  & (deg) \\ 
\hline
J0114+0722 & 0.84$\pm$0.04 & 2.13$\pm$0.24 & 0.23$\pm$0.02 &  73.3$\pm$6.2 \\ 
J0256+0153 & 1.16$\pm$0.06 & 1.35$\pm$0.21 & 0.33$\pm$0.03 &  72.5$\pm$4.3 \\ 
J0737+4825 & 1.08$\pm$0.35 & [4.00]        & 0.14$\pm$0.12 & $-$70.0$\pm$32.6 \\ 
J0821+4542 & 0.14$\pm$0.02 & [4.00]        & [0.00]        &  [0.0] \\ 
J0921+2854 & 0.70$\pm$0.01 & 0.98$\pm$0.06 & 0.10$\pm$0.02 &  36.7$\pm$9.6 \\ 
J1254+1857 & 0.73$\pm$0.14 & [4.00]        & 0.38$\pm$0.12 & $-$16.1$\pm$14.5 \\ 
J1330+3800 & 0.87$\pm$0.34 & [4.00]        & 0.75$\pm$0.16 & $-$16.3$\pm$9.7 \\ 
J1412+5204 & 0.61$\pm$0.15 & [4.00]        & 0.58$\pm$0.08 &  15.5$\pm$7.4 \\ 
J1442+4055 & 0.99$\pm$0.10 & 4.19$\pm$0.70 & 0.21$\pm$0.03 &  68.2$\pm$6.1 \\ 
J1452+4224 & 1.09$\pm$0.01 & [4.00]        & 0.30$\pm$0.01 & $-$23.8$\pm$1.1 \\ 
J1458-0202 & 2.13$\pm$0.24 & 3.81$\pm$0.30 & 0.36$\pm$0.02 &  30.6$\pm$1.6 \\ 
J1537+3014 & 3.57$\pm$2.09 & 5.90$\pm$1.92 & 0.37$\pm$0.04 &  87.2$\pm$5.1 \\ 
J2146-0047 & 0.38$\pm$0.28 & [4.00]        & [0.00]        &  [0.0]\\ 
\hline
\end{tabular}
\end{center}
The parameters $r_e$, $n$, $e$, and $\theta_e$ denote the effective
radius, Sersic index, ellipticity, and the position angle (measured
counter-clockwise from North where North is up at 0 deg), respectively. See
Table~\ref{tab:posmag} for the magnitudes.
\end{table}

\subsection{Notes on interesting individual systems}
\label{sec:individual}

Here, we discuss the properties of some of the interesting systems from the BQLS.

\subsubsection{SDSS~J0818+0601}
The spectrum shown in Figure~\ref{fig:spec} indicates that the two stellar
components separated by $\Delta\theta\sim 1''$ have similar spectra
($\Delta V=226$~km~s$^{-1}$), including broad absorption features
in the Ly$\alpha$ and CIV emission lines. However, no lens galaxy was
found in our SOAR images (420~sec in $i$). If this is
a gravitational lens, the non-detection of the lens suggests it is
at a relatively high redshift ($z_l\ga 1$).

\subsubsection{SDSS~J0821+0735}
While the spectra of the components (Figure~\ref{fig:spec}) indicate
that the quasar redshifts are almost same ($\Delta
V=319$~km~s$^{-1}$), their 
SEDs appear to be slightly different. The emission line
features have different profiles and the continuum shows differences in
the slope. Some differences in the SEDs of lensed counterparts can be
attributed to dust extinction in the lens galaxy
\citep[e.g.,][]{Falco1999} or micro-lensing. However, the UH88 imaging
observation (1440~sec in $I$) also failed to detect any lens
between the quasar components. 

\subsubsection{SDSS~J0921+2854}
After we confirmed the lensing nature of this object, an archival search
revealed that this lens was independently discovered by E. Ofek et al., and was
observed by the {\it Hubble Space Telescope} Cycle 20 program GO-13001
although their results are unpublished. There
are also X-ray and radio detections for this system, probably emanating from
the lensed quasar.

\subsubsection{SDSS~J0930+4614}
This is a quasar pair at the same redshift ($\Delta
V=111$~km~s$^{-1}$) but with quite different
SEDs (Figure~\ref{fig:spec}). The fainter component has broad
absorption lines associated with Ly$\alpha$, SiIV, and CIV emission
lines, while the brighter component has no broad absorption features. Given
the markedly different SEDs, this is likely to be a physical quasar
pair rather than a lens. 

\subsubsection{SDSS~J1043+4320}
This is a quasar pair with similar redshifts ($\Delta
V=1778$~km~s$^{-1}$).  There are strong absorption lines in the
CIV emission lines of both components, but the overall SED shapes appear
to be different (see Figure~\ref{fig:spec}).  It is important to conduct
a deep imaging to search for a possible lens galaxy.

\subsubsection{SDSS~J1124+5710}
The SEDs of this quasar pair appear similar but have slightly different
redshifts ($\Delta V=496$~km~s$^{-1}$, Figure~\ref{fig:spec}).
There is no indication of a lens galaxy in our UH88 imaging (1440~sec in
$I$). Hence, we conclude that this is likely a binary quasar system. 

\subsubsection{SDSS~J1309+5617}
While the redshifts of the two components are almost identical ($\Delta
V=169$~km~s$^{-1}$), the significantly different shapes of the
Ly$\alpha$ and CIV emission lines mean that this system is likely to be a
binary quasar system rather than a lens. 

\subsubsection{SDSS~J1405+1350}
The SED of the fainter quasar has some similarity with the SED of the
brighter quasar but the emission features have very low signal-to-noise
ratios. The redshifts are not identical but imply that the quasars are
probably physically associated ($\Delta V=1400$~km~s$^{-1}$).
Additionally, the non-detection of a putative lens galaxy suggests this
might be a binary quasar system.

\subsubsection{SDSS~J1452+4224}
At $z_s\approx 4.8$\footnote{Redshifts from Ly$\alpha$ at such high redshifts
are strongly affected by the Ly$\alpha$ forest \citep{McGreer2010}.
Therefore, we give fewer significant digits on the redshift estimate.},
the quasar redshift of this lens is one of the highest redshifts known
and is comparable to the lens SDSS~J0946+1835 \citep{McGreer2010}.  The total
lensing magnification is predicted to be $\sim 8$ by our mass models
in Section~\ref{sec:massmod}. The lens galaxy is quite visible
in the spectrum and the image, and has one of the lowest measured lens redshift
($z_l=0.382$) in our sample.  Coincidentally,
the lens galaxy of SDSS~J0946+1835 is at a similar redshift but more
massive given the wider image separations of the lensed quasars.

\subsubsection{SDSS~J2146-0047}
\label{sec:sdssj2146}
While the lensing nature of this object was confirmed by our imaging and
spectroscopic follow-up observations, \citet{Agnello2015} recently
reported an independent discovery of this lens using Dark Energy Survey
data. They tentatively assigned a lens redshift of $z_l=0.799$ based on
MgII and FeII absorption features in the quasar spectra. The absorption
feature is also seen in the Subaru spectrum shown in
Figure~\ref{fig:spec}. 

\section{Mass modelling}
\label{sec:massmod}

\begin{figure*}
\begin{center}
\includegraphics[scale=0.90]{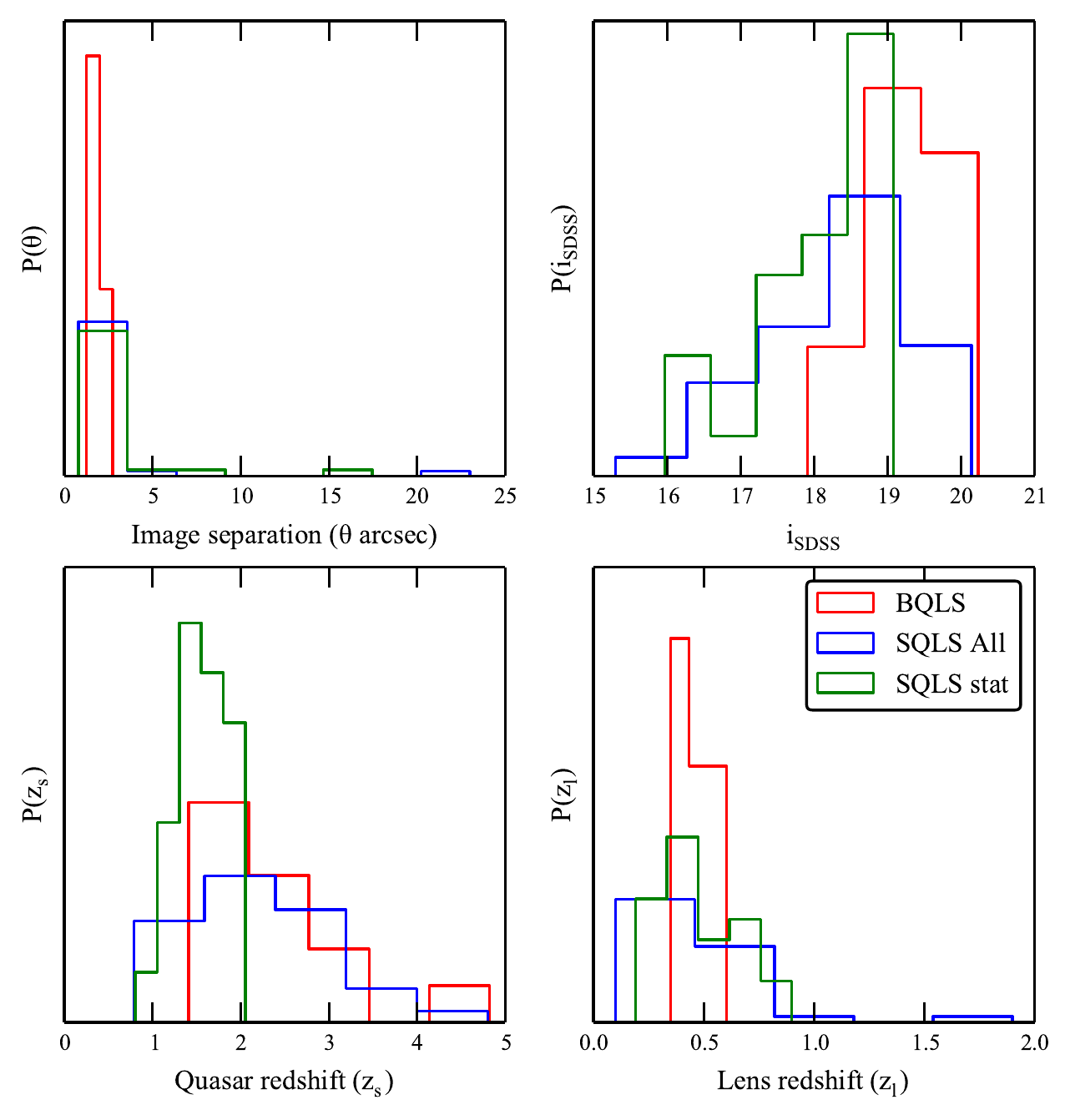}
\caption{ \label{fig:comp} 
Comparison of the BQLS with the SQLS ``statistical'' and ``all''
samples. Owing to the smaller sample size of BQLS, a qualitative
comparison is more meaningful. Each histogram is thus normalised and
integrates to unity. Lenses with no known lens redshifts are excluded
from the panel showing the redshift distributions. BQLS lenses are
fainter and have smaller image separations. The quasar redshift
distribution and the peak of the lens redshift distribution of the BQLS
sample are similar to the SQLS ``all'' sample. }
\end{center}
\end{figure*}

We created mass models for the 13 spectroscopically confirmed lens
systems with the modelling software, {\sc glafic} \citep{Oguri2010}.
The redshifts of the quasars and the lens galaxies were taken from
Table~\ref{tab:spec}. For those systems with no lens redshifts, we
adopted a fiducial value of $z_l=0.5$ ($\sim$ peak of the redshift
distribution of typical lenses e.g., see Fig~\ref{fig:comp}), as a lens
redshift is required in the estimate of the velocity dispersion. 

We assumed that the mass distribution of the lens galaxies is
isothermal. More specifically, we either used the singular isothermal sphere (SIS) or
ellipsoid (SIE) model for the lenses.  Parameters such as the mass,
ellipticity and the position angle ($\theta_e$) of the lens model and
the true (unlensed) position of the lensed quasar are allowed to be
free. The positions and flux ratios\footnote{The flux ratios are often
known to be affected by dust, substructure, intrinsic quasar variability
and micro-lensing. Thus, a simple globally smooth models such as the ones
tested here may fail to fit the flux ratios and cannot necessarily
be used to rule out the lens hypothesis.} of the lensed quasar images served as
constraints. We always begin with the SIS model and add further
complexity such as adding more parameters only if the SIS
results in a poor fit. The best fit parameters from the mass models are
reported in Table~\ref{tab:massmod}. 

For SDSS~J1330+3800, neither the SIS nor the SIE model could fit the
constraints well. Hence, for this system, we used an SIS model with the lens
galaxy position allowed to vary within the uncertainties from the {\sc
galfit} model. In a few other cases, we also used priors on the position
angle of the lens based on the photometric models (see
Table~\ref{tab:massmod}). In general, we could use the {\sc galfit}
errors on the positions and fluxes of the lensed images but had to relax
them\footnote{This is reasonable because the errors from {\sc galfit}
are known to be an underestimate \citep[e.g.,][]{Rusu2015}} in a few
cases in order to find a reasonable $\chi^2$. The errors were relaxed up
to $\sim 0\farcs15$ and were always smaller than the pixel size.
The best fit Einstein radii (or the velocity dispersions) of the BQLS
lenses, which is a proxy for the mass of the lens galaxy, are similar
to typical galaxy-scale lenses \citep[e.g.][]{Bolton2006b}. 

\begin{table}
\begin{center}
\caption{\label{tab:massmod} Best fit lens mass model parameters. }

\begin{tabular}{clrr}
\hline
Object & $\theta_{\rm Ein}$ & $e$ & $\theta_e$ \\ 
ID & (\arcsec) &  & (deg) \\ 
\hline
SDSS~J0114+0722 & 0.83 &  0.42    &   45    \\
SDSS~J0256+0153 & 1.03 &  0.74    &   42    \\
SDSS~J0737+4825 & 0.81 &  0.25    &  [$-$82]   \\
SDSS~J0821+4542 & 0.63 &  0.51    &  $-$21    \\
SDSS~J0921+2854 & 0.96 &  0.53    &   28      \\
SDSS~J1254+1857 & 1.15 &   --     &   --    \\
SDSS~J1330+3800 & 0.89 & [$-$0.553] &  [0.321]$\dagger$  \\
SDSS~J1412+5204 & 0.84 &  0.45    &  52     \\
SDSS~J1442+4055 & 1.03 &  --      &  --     \\
SDSS~J1452+4224 & 0.87 &  --      &  --     \\
SDSS~J1458-0202 & 1.15 &  0.27    &  14     \\
SDSS~J1537+3014 & 1.14 &  0.21    & [88]     \\
SDSS~J2146-0047 & 0.71 &  --      &  --     \\
\hline
\end{tabular}
\flushleft{ The parameter $\theta_{\rm Ein}$, $e$, and $\theta_e$ denote
the Einstein radius, ellipticity, and the position angle of the mass
distribution, respectively. The numbers in bracket indicate which priors
were used from the {\sc galfit} model (see Table~\ref{tab:posmag}).
$\dagger$ - This is an SIS model with priors on the lens position. Instead
of the ellipticity and position angle, the last two columns list the
best fit offset (in arcseconds) of the lens from one of the lensed
quasars in x and y direction, respectively. We use the standard
cartesian convention where x and y are positive to the right and up,
respectively.}
\end{center}
\end{table}

\begin{table}
\begin{center}
\caption{\label{tab:lens} Summary of the confirmed quasar lenses. }
\begin{tabular}{ccrcc}
\hline
Object ID & $z_s$ & $z_l$ & $\Delta\theta(\arcsec)$ & $i$ \\
\hline
SDSS~J0114+0722 & 1.828 & 0.408 & 1.70 & 19.1 \\
SDSS~J0256+0153 & 2.600 & 0.603 & 1.93 & 20.0 \\
SDSS~J0737+4825 & 2.892 & --   & 1.54 & 18.5 \\
SDSS~J0821+4542 & 2.066 & 0.349 & 1.35 & 18.9 \\
SDSS~J0921+2854 & 1.410 & 0.445 & 1.91 & 18.9 \\
SDSS~J1254+1857 & 1.717 & 0.555 & 2.32 & 19.5 \\
SDSS~J1330+3800 & 2.254 & --    & 1.44 & 20.1 \\
SDSS~J1412+5204 & 2.952 & --    & 1.53 & 19.0 \\
SDSS~J1442+4055 & 2.575 & --    & 2.13 & 17.9 \\
SDSS~J1452+4224 & 4.819 & 0.382 & 1.59 & 19.1 \\
SDSS~J1458-0202 & 1.724 & --    & 2.15 & 19.1 \\
SDSS~J1537+3014 & 1.553 & 0.490 & 2.30 & 19.6 \\
SDSS~J2146-0047 & 2.381 & 0.799 &  1.39 & 20.2 \\
\hline
\end{tabular}
\flushleft{ 
The image separations $\Delta\theta$ are computed from 
Table~\ref{tab:posmag} and the $i$-band PSF magnitudes are from the SDSS database.}
\end{center}
\end{table}

\begin{table*}
\begin{center}
\caption{\label{tab:pair} Summary of confirmed quasar pairs.}
\begin{tabular}{ccccccccc}
\hline
Object ID & RA$_{\rm B}$ & Dec$_{\rm B}$ & $z_{\rm A}$  & $z_{\rm B}$ & $\Delta\theta(\arcsec)$ & $i_{\rm A}$ & $i_{\rm B}$ & $\Delta$~V (km~s$^{-1}$)\\
\hline
SDSS~J0213-0421 & 02:13:22.865 & $-$04:21:34.341  & 1.911  &  0.992   &  2.0  &   20.8  & 18.8  &  $>2000$  \\
SDSS~J0818+0601 & 08:18:30.420 &   +06:01:37.860  & 2.359  &  2.361   &  1.1  &   18.1  & 20.0  &  226  \\
SDSS~J0821+0735 & 08:21:43.241 &   +07:35:45.108  & 2.383  &  2.387   &  1.3  &   20.7  & 21.6  &  319  \\
SDSS~J0928+4332 & 09:28:39.048 &   +43:32:42.141  & 3.694  &  2.995   &  1.6  &   21.1  & 20.3  &  $>2000$   \\
SDSS~J0930+4614 & 09:30:20.986 &   +46:14:23.260  & 2.393  &  2.394   &  1.5  &   18.4  & 19.9  &  111  \\
SDSS~J1043+4320 & 10:43:25.025 &   +43:20:48.958  & 2.245  &  2.225   &  1.7  &   21.4  & 20.7  &  1778  \\
SDSS~J1124+5710 & 11:24:55.440 &   +57:10:58.120  & 2.309  &  2.315   &  2.0  &   18.6  & 19.7  &  496  \\
SDSS~J1309+5617 & 13:09:27.338 &   +56:17:41.161  & 2.513  &  2.515   &  2.8  &   20.5  & 21.6  &  169  \\
SDSS~J1405+1350 & 14:05:56.865 &   +13:50:39.827  & 2.345  &  2.361   &  1.7  &   21.7  & 22.0  &  1400  \\
SDSS~J1548+2830 & 15:48:50.776 &   +28:30:12.714  & 3.208  &  1.487   &  1.7  &   20.8  & 20.9  &  $>2000$   \\
SDSS~J2245+2548 & 22:45:55.801 &   +25:48:33.548  & 2.995  &  2.178   &  2.0  &   20.4  & 20.7  &  $>2000$   \\
\hline
\end{tabular}
\end{center}
See Table~\ref{tab:spec} for the RA and Dec of the brighter quasar image
(A). The positions and the $i$-band magnitudes in this table are from
the SDSS database except for J0818+0601 where we used the
astrometrically calibrated SOAR images. 
\end{table*}

\section{Discussion}
\label{sec:disc}

\subsection{Properties of the BQLS}
The basic properties of the 13 quasar lenses are summarized in
Table~\ref{tab:lens}. We also compare the image separation, magnitude,
and redshift distributions of the newly discovered quasar lens systems
with those of the SQLS sample in Figure~\ref{fig:comp}. In the redshift
comparison, we exclude systems for which the lens redshifts are not
known. There are two SQLS lens samples \citep{Inada2012}: i) ``all'' -
the entire lens sample which includes some heterogeneously selected
lenses and ii) ``stat'' - this is a statistically well-defined sample
with high completeness. The SQLS statistical lens
sample is limited to the quasar redshift range of $0.6<z<2.2$, $i$-band
magnitude $i<19.1$, image separations  $1\arcsec < \Delta \theta <
30\arcsec$, and flux ratios for doubles larger than $10^{-0.5}$, in
order to accurately characterize the selection function. 

Since the BOSS sample size is much smaller, we show normalised distributions to
facilitate qualitative comparison of the properties of these samples. The image
separation distribution suggests that the BOSS sample finds a larger fraction
of small image separation ($\Delta\theta\la 2''$) lenses compared to the
SQLS sample. One likely reason for this difference is that we preferentially
selected small-separation lens candidates for follow-up confirmation because 
larger separation quasar pairs are more likely to be physical
pairs rather than real lenses \citep[e.g.,][]{Kochanek1999, Hennawi2006b}. Another
possibility may be fiber collisions. Large-separation lenses are produced by 
clusters of galaxies typically at $z\sim 0.5$, and therefore there are usually
many member galaxies around the quasar images that are selected as Constant
MASS (CMASS) galaxies for spectroscopy. This may reduce the chance of these
quasar images being spectroscopically observed by BOSS. As mentioned earlier,
this was indeed the case for SDSS~J2222+2745 where none of the quasar images
were observed by BOSS while spectra were taken of several nearby
member galaxies of this cluster lens.

We find that 7 out of the 13 confirmed quasar lens systems are located
at $z>2.2$. The fraction of high-redshift quasar lenses is not
particularly high compared with the ``all'' SQLS sample (see
Figure~\ref{fig:comp}). This result may be partly explained by the fact
that some of the new lens systems were originally targeted as CMASS
galaxies rather than quasars because of the dominant lens galaxy
component. Indeed, for many of these cases, the lens galaxies are
visible in the SDSS spectra which then provide the lens redshifts for these
systems (see Table~\ref{tab:spec}). Spectroscopic quasars selected in
this manner are not necessarily high-redshift quasars. This also
explains the peak of the lens redshift distribution at $z_l\sim 0.5$,
which is close to the median redshift of the BOSS CMASS galaxy sample. 

Most of the BQLS lenses are fainter because BOSS target selection
selects fainter quasars. The BOSS DR12 quasar catalogue contains 25
previously known lenses, many found by the SQLS.  About two-thirds of
this sample are bright lens systems ($i<18$) with quasars at low
redshifts ($z<2.2$). We excluded this known sample before the visual
selection of the BQLS lens candidates.

A sub-sample of the BOSS quasars are selected homogeneously and have a
well-defined selection function which is called the CORE sample
\citep{Ross2012}. Lenses present in the CORE sample could be used to
define a statistical sample for the BQLS provided the visual selection
process is calibrated. However, only two out of the 13 confirmed lenses
are in the CORE sample (using the boss\_target1 flag for
QSO\_CORE\_MAIN).  It is unlikely that our visual inspection process is
highly inefficient, so it is unlikely that we are missing a large number
of true lenses. Therefore, even if we were to detect most of the true
lenses, a statistically well-defined sample derived from this parent
sample will still be probably too small for any statistical study. The ``point-source''
requirement for the CORE sample is probably the cause of this low yield.

\subsection{Quad fraction}

All of the 13 confirmed quasar lenses are two-image quasar lens systems.
At galaxy scales, we expect about 15$-$20\% of the lenses to be
quadruply imaged for the BOSS quasar lens sample with $i\la 20-21$
\citep{Oguri2010b} which means we should expect two or more quad lenses
in our sample. There are several possible explanations for the lack of
quad lenses in the BOSS quasar lens sample. First of all, the BOSS
quasar target selection selects only point sources to increase its
efficiency \citep{Ross2012}, whereas small-separation quad lens systems
are often classified as extended sources in the SDSS \citep{Oguri2006}.
Indeed, all the four small-separation ($\Delta\theta<2''$) quad lenses
discovered in the SQLS \citep[see][]{Inada2012} are found to be
classified as extended sources in the SDSS dataset. In addition,
the visual inspection method might be biased against selecting quad
lenses because the quads appear to have odd morphology. And,
spectroscopic follow-up observations might be biased because
single long-slit mode is used to target the double quasar candidate images along
with the candidate lens galaxy for confirmation. Therefore it is of great
interest to apply more sophisticated algorithms
\citep[e.g.,][]{Chan2015,Agnello2015b} to the quasars in BOSS to search
for quad lenses that are missing in the current lens sample.  

\subsection{Quasar pairs}
\label{sec:qpair}

Our follow-up spectroscopy identified an additional 11 quasar pairs with
small angular separations of $\Delta\theta<3''$ (see
Table~\ref{tab:pair} for a summary). We expect most of these to be
either projected pairs or binary quasars, based on their different SEDs
and the lack of an obvious lens. Among the 11 quasar pairs, 7 pairs
have the velocity difference $\Delta$V smaller than 2000~km~s$^{-1}$ and
therefore satisfy the criterion of physical binaries used in
\citet{Hennawi2006b}. The velocity difference may also be useful for
distinguishing binary quasars from lensed quasars. We find that the
typical $\Delta$V of the confirmed lenses measured from our follow-up
spectra is $<40$~km~s$^{-1}$ and $\sim 100$~km~s$^{-1}$ in a few cases,
which can be regarded as a typical error on our velocity difference
measurements. From Table~\ref{tab:pair}, we find that most of the quasar pairs
have velocity differences significantly larger than those of the confirmed lenses,
indicating that they are less likely to be lensed images of a single
source but rather are physically associated distinct quasars. 

In general, there are still only a small number of quasar pairs known
\citep{Hennawi2006,Hennawi2010,Myers2008,Kayo2012}, and our sample adds
significantly to the existing pair samples. For comparison, the SQLS
only identified 8 binary quasars with $\Delta\theta<3''$ \citep[see][for
a compilation]{Kayo2012}, which is comparable to the number of new
binary quasars from the BQLS. The small-separation binary quasars have
been used to study the very small scale clustering of quasars to discuss
the possible role of mergers in enhancing quasar activities
\citep{Hennawi2006,Hopkins2008,Myers2008,Shen2010,Kayo2012}.  Projected
quasar pairs are also useful for studying the distribution of absorbers
around quasars \citep{Hennawi2006b}.  

\section{Summary and Conclusion}
\label{sec:conc}

We present the initial results from BQLS, a systematic search for lensed
quasars in the SDSS-III/BOSS data. We applied the same technique that we
used for finding lensed quasars in the SDSS DR7 (SQLS).  Here, we
report the discovery of 13 confirmed quasar lenses. In
addition, we present the discovery of 11 quasar pairs, some of which may
consist of physical binaries.  This sub-sample may still contain a few
unrecognised lenses.  The sample of new lenses includes one of the highest
redshift lensed quasar (SDSS~J1452+4224, $z\approx 4.8$) found to date.
SDSS~J1442+4055 was also discovered independently by
\citet{Sergeyev2016}. All
the confirmed lenses from the BQLS have only two images. The lack of
quad lenses is probably because BOSS only selects point-like quasars.
We note that our follow-up observations are still incomplete with an
additional $\sim 50$ good lens candidates that need verification.
However, this sample is likely to have fewer real lenses since we
conducted the follow-up observations of the most promising candidates first.

Compared to the SQLS, the selection function of the BQLS sample is
complicated owing to the complex selection of the initial BOSS quasar
targets. In addition to the incomplete follow-up observations, 
a complex selection function prevents us from using
the BQLS sample for statistical studies of either the lens population or
cosmology \citep[e.g.,][]{Oguri2012}. The CORE sample from BOSS, which is
a uniform and well-defined quasar sample could be used to produce a
statistical BQLS sample, yielded only two lenses. This is probably
because the quasars are required to be point-like in the CORE sample. 

A qualitative comparison of the initial BQLS sample with the SQLS shows
that BQLS lenses are somewhat fainter and have smaller image
separations, presumably due to the BOSS selection criteria. Furthermore,
the BQLS and SQLS lensed quasars have similar redshift distributions in
spite of the higher redshift selection used for BOSS quasars. This is
because many of the lenses in our sample were photometrically identified
as galaxies before they were spectroscopically confirmed by BOSS to be
quasars. In addition, about half of the lensed quasars have redshifts
lower than the quasar redshift range targeted by BOSS. These lenses
would have been missed in the absence of BOSS CMASS galaxy
spectroscopy. 

This study  provides useful guidance for ongoing quasar lens surveys in
Hyper
Suprime-Cam\footnote{{http://www.naoj.org/Projects/HSC/surveyplan.html}}
and Dark Energy Survey \citep{TheDarkEnergySurveyCollaboration2005} in which
even fainter quasar lenses will be discovered.

\section*{Acknowledgments}
The work of MO and AM was supported in part by World Premier
International Research Center Initiative (WPI Initiative), MEXT, Japan.
This work was also supported by Grant-in-Aid for Scientific Research
from the JSPS (26800093 and 24740171). 
AM would like to thank S. More and J. Silverman for useful suggestions.
AM acknowledges the support of the Japan Society for Promotion of
Science (JSPS) fellowship.
A.M.M. acknowledges the support of NSF grant AST-1211146.
The authors would like to thank the anonymous referee for useful
suggestions that improved the paper.

Funding for SDSS-III has been provided by the Alfred P. Sloan
Foundation, the Participating Institutions, the National Science
Foundation, and the U.S. Department of Energy Office of Science. The
SDSS-III web site is {http://www.sdss3.org/}. SDSS-III is managed by the Astrophysical Research Consortium for the Participating Institutions of the SDSS-III Collaboration including the University of Arizona, the Brazilian Participation Group, Brookhaven National Laboratory, Carnegie Mellon University, University of Florida, the French Participation Group, the German Participation Group, Harvard University, the Instituto de Astrofisica de Canarias, the Michigan State/Notre Dame/JINA Participation Group, Johns Hopkins University, Lawrence Berkeley National Laboratory, Max Planck Institute for Astrophysics, Max Planck Institute for Extraterrestrial Physics, New Mexico State University, New York University, Ohio State University, Pennsylvania State University, University of Portsmouth, Princeton University, the Spanish Participation Group, University of Tokyo, University of Utah, Vanderbilt University, University of Virginia, University of Washington, and Yale University. 

Based in part on data collected at Subaru Telescope, which is operated by the National Astronomical Observatory of Japan.
Use of the UH 2.2-m telescope for the observations is supported by NAOJ.
Some of the data presented herein were obtained at the W.M. Keck Observatory, which is operated as a scientific partnership among the California Institute of Technology, the University of California and the National Aeronautics and Space Administration. The Observatory was made possible by the generous financial support of the W.M. Keck Foundation.
The authors wish to recognize and acknowledge the very significant cultural role and reverence that the summit of Mauna Kea has always had within the indigenous Hawaiian community.  We are most fortunate to have the opportunity to conduct observations from this mountain.
A part of this work is based on observations obtained at the MDM Observatory,
operated by Dartmouth College, Columbia University, Ohio State
University, Ohio University, and the University of Michigan.
Based on observations obtained at the Southern Astrophysical Research (SOAR) telescope, which is a joint project of the Minist\'{e}rio da Ci\^{e}ncia, Tecnologia, e Inova\c{c}\~{a}o (MCTI) da Rep\'{u}blica Federativa do Brasil, the U.S. National Optical Astronomy Observatory (NOAO), the University of North Carolina at Chapel Hill (UNC), and Michigan State University (MSU).

\bibliographystyle{apj}
\bibliography{references_papers}

\end{document}